\documentclass[fleqn,usenatbib]{mnras}

\usepackage{newtxtext} 
\usepackage{amsmath}
\usepackage{makecell}
\usepackage{booktabs}
\usepackage{pdflscape}

\usepackage[T1]{fontenc}
\usepackage{CJK}

\DeclareRobustCommand{\VAN}[3]{#2}
\let\VANthebibliography\thebibliography
\def\thebibliography{\DeclareRobustCommand{\VAN}[3]{##3}\VANthebibliography}

\usepackage{graphicx}	
\usepackage{amsmath}	
\usepackage{amssymb}	

\def\eg{{ e.g.,\ }}
\def\ie{{ i.e.,\ }}
\def\deg{^\circ}
 
\def\kpc{{\rm\,kpc}}
\def\Mpc{{\rm\,Mpc}}
\def\pc{{\rm\,pc}}
\def\degx{{\rm\,deg}}
\def\msun{{\rm\,M$_{\odot}$}}
\def\FeH{{\rm[Fe/H]}}

\title[Dwarf Galaxy Detection with CSST]{Local Group Dwarf Galaxy Detection Limit in the CSST survey}

\begin{document}
\begin{CJK*}{UTF8}{gkai}  
\author[Han Qu et al.]{
Han Qu (曲涵)$^{1,2}$\thanks{E-mail: qvhan@pmo.ac.cn}
Zhen Yuan (袁珍)$^{3}$\thanks{corresponding author, E-mail: zhen.yuan@astro.unistra.fr}
Amandine Doliva-Dolinsky,$^{3}$
Nicolas F. Martin$^{3,4}$,
\newauthor 
Xi Kang (康熙)$^{5,1}$, 
Chengliang Wei (韦成亮)$^{1}$, 
Guoliang Li (李国亮)$^{1,2}$,
Yu Luo (罗煜)$^{1}$, 
\newauthor 
Jiang Chang (常江)$^{1}$, 
Chaowei Tsai (蔡肇伟)$^{6,7,8}$, 
Zhou Fan (范舟)$^{6}$,  
Rodrigo Ibata$^{3}$
\\
$^{1}$Purple Mountain Observatory, Chinese Academy of Sciences, Nanjing 210008, China\\
$^{2}$School of Astronomy and Space Sciences, University of Science and Technology of China, Hefei 230026, China\\
$^{3}$Université de Strasbourg, CNRS, Observatoire Astronomique de Strasbourg, UMR 7550, F-67000 Strasbourg, France\\
$^{4}$Max-Planck-Institut f\"ur Astronomie, K\"onigstuhl 17, D-69117, Heidelberg, Germany \\
$^{5}$Zhejiang University-Purple Mountain Observatory Joint Research Center for Astronomy, Zhejiang University, Hangzhou 310027, China\\
$^{6}$National Astronomical Observatories, Chinese Academy of Sciences, 20A Datun Road, Chaoyang District, Beijing 100012, China\\
$^{7}$Institute for Frontiers in Astronomy and Astrophysics, Beijing Normal University,  Beijing 102206, China\\
$^{8}$University of Chinese Academy of Sciences, Beijing 100049, China
}

\date{Accepted 2023 April 27. Received 2023 March 30; in original form 2022 December 20 }

\pubyear{2023}

\label{firstpage}
\pagerange{\pageref{firstpage}--\pageref{lastpage}}
\maketitle
\begin{abstract}

We predict the dwarf galaxy detection limits for the upcoming Chinese Space Station Telescope (CSST) survey that will cover 17,500 $\degx^{2}$ of the sky with a wide field of view of 1.1 deg$^2$. The point-source depth reaches 26.3 mag in the $g$ band and 25.9 mag in the $i$ band. Constructing mock survey data based on the designed photometric bands, we estimate the recovery rate of artificial dwarf galaxies from mock point-source photometric catalogues. The detection of these artificial dwarf galaxies is strongly dependent on their distance, magnitude and size, in agreement with searches in current surveys. We expect CSST to enable the detection of dwarf galaxies with $M_V = -3.0$ and $\mu_{250} = 32.0$ mag/arcsec$^2$ (surface-brightness limit for a system of half-light radius $r_{\rm h}$ = 250 $\pc$) at $400 \kpc$,  and $M_V = -4.9$ and $\mu_{250} = 30.5$ mag/arcsec$^2$ around the Andromeda galaxy. Beyond the Local Group, the CSST survey will achieve $M_V = -5.8$, and $\mu_{250}$ = 29.7 mag/arcsec$^2$ in the distance range of 1--2 Mpc, opening up an exciting discovery space for faint field dwarf galaxies. With its optical bands, wide survey footprint, and space resolution, CSST will undoubtedly expand our knowledge of low-mass dwarf galaxies to an unprecedented volume.

\end{abstract}
\end{CJK*}




\section{Introduction}
Within the paradigm of hierarchical structure formation, massive galaxies grew through the absorption of low-mass systems, most of which are dwarf galaxies \citep{Lacey1993, Vogelsberger2014}. In the Local Group, some of these building blocks survive until now and become satellites of the Milky Way (MW) and the Andromeda galaxy (M31). The number of these dwarf galaxies can provide tests for cosmological models, as well as constraints on baryonic physics that governs galaxy evolution \citep{Bovill2009, Bovill2011, Phillips2015,Bullock2017}. 

Significant progress has been made in searching dwarf satellite galaxies around the MW-M31 pair in recent years \citep{Walsh2007, Walsh2009, Martin2013a, Bechtol2015, Drlica-Wagner2015, Koposov2015, Laevens2015a, Laevens2015b,Simon2019}. Up to now,  nearly a hundred dwarf galaxies have been detected in the Local Group: 59 in the MW with $M_{V} < -0.8 \pm 0.9$ and 39 in M31 with $M_{V} < -5.9 \pm 0.7$ thanks to the large scale digital survey including the Sloan Digital Sky Survey \citep[SDSS;][]{Abazajian2003}, the Dark Energy Survey \citep[DES;][]{Abbott2018}, the Pan-STARRS survey \citep{Chambers2016}, or the Pan-Andromeda Archaeological Survey \citep{McConnachie2018}.

In the era of optical and near-infrared wide surveys constructed with space telescopes (\eg Euclid, \citealt{Wallner2017}; Roman, \citealt{Spergel2015}), the Chinese Space Station Telescope (CSST) survey is planned to perform both photometric imaging and slitless spectroscopic observations and will cover 17,500 $\degx^{2}$ in its planned ten years of operation \citep{Zhan2011,Cao2018,Gong2019}. The CSST is a 2-meter space telescope and shares the same orbit as the Chinese Space Station, which will be launched in the next few years. The main survey camera module has a large field of view (1.1\,$\rm deg^{2}$), a high spatial resolution ($\sim$0.15''), and is equipped with a set of near-ultra-violet to near-infrared filters ($NUV$ and $ugrizy$). The instrumental design of the CSST makes it an optimal tool to search for low-mass/faint dwarf galaxies within the Local Volume. In this paper, we make the first predictions of the detectability of dwarf galaxies within $2\Mpc$ from mock photometric catalogues.

The paper is structured as follows. In Sec.~\ref{sec:mock data}, we estimate the completeness and photometric uncertainties of point-source detections with CSST. Based on that, we construct a series of mock catalogues by injecting artificial dwarf galaxies in the modeled contamination from MW foreground stars. The dwarf galaxy search algorithm is explained in Sec.~\ref{sec:method}. The results from applying this algorithm to the mock catalogs are discussed in Sec.~\ref{sec:appli}, in which we compare the detection limits of the CSST survey with limits from existing surveys and the known Local Group dwarf galaxies. Sec.~\ref{sec:sum} closes the paper with a summary and discussion.

\begin{figure}
   \centering
   \includegraphics[width=8.5cm, angle=0]{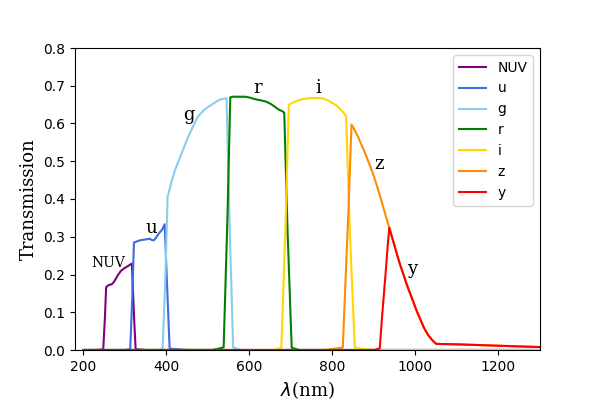}     
   \caption{Transmission curves of the 7 CSST filters, covering wavelengths from the near ultraviolet, to the near infrared ($NUV, u, g, r, i, z$, and $y$ bands). 
   }
   \label{filter}
\end{figure}

\begin{table*} 
	\caption{Designed Parameters of CSST}
	\setlength{\tabcolsep}{62pt}
	\label{tab:CSST}
	\centering  
	\begin{tabular}{c|c} 
		\hline
		\hline
		\multicolumn{2}{c}{Survey Characteristics}\\
		\hline
		Wide survey   & 15000 deg$^{2}$ (|b|>20$\deg$) + 2500 deg$^{2}$ (15$\deg$<|b|<20$\deg$)\\
		Deep survey   & 400 deg$^{2}$ (selected areas) \\
		\hline
	\end{tabular}
	\setlength{\tabcolsep}{10pt}
	\begin{tabular}{c|ccccccc } 
    \multicolumn{8}{c}{Photometric System}\\
	\hline  
		Band & $NUV$ & $u$ & $g$ & $r$& $i$ & $z$ & $y$\\  
		\hline  
        Wavelength ($\lambda_{-90}-\lambda_{+90}$ nm) & 255-317 & 322-396 & 403-545& 554-684& 695-833 & 846-1065 & 937-1065\\ 
		Wide imaging (5$\sigma$ depth)&25.4&25.4&26.3&26.0&25.9&25.2&24.4\\
		Deep imaging (5$\sigma$ depth)&26.7&26.7&27.5&27.2&27.0&26.4&25.7\\
		\hline
	\end{tabular}
\end{table*}

\begin{figure*}
    \centering
	\includegraphics[width=19.0cm,angle=0]{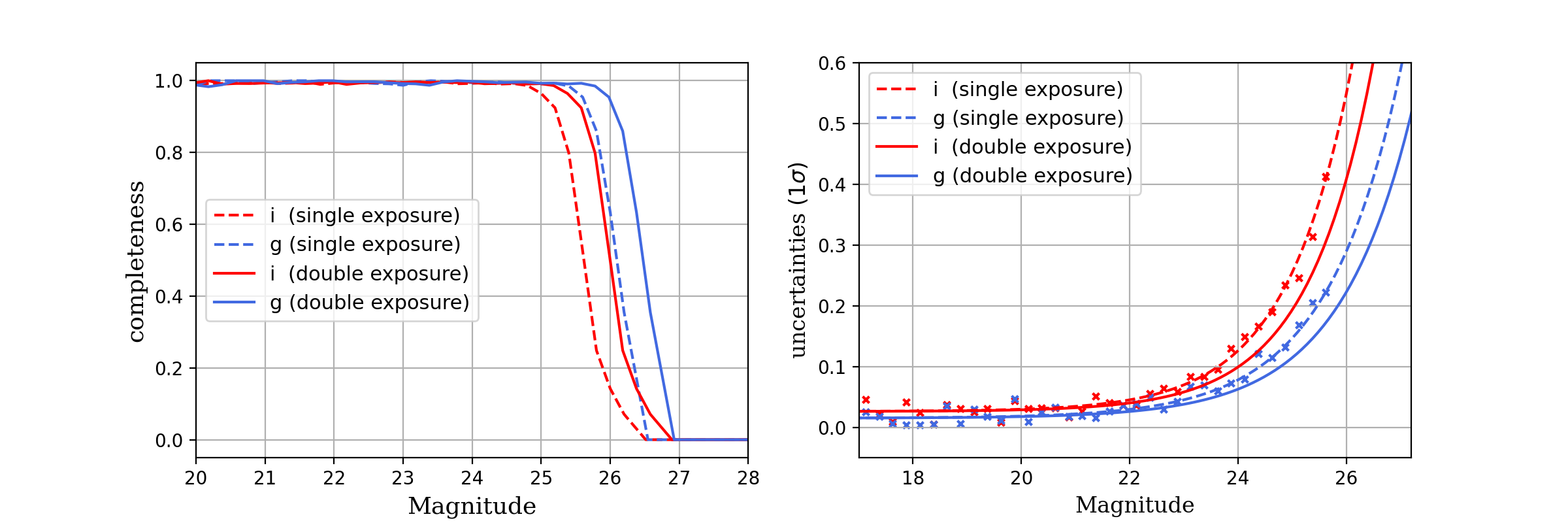}
    \caption{\emph{Left:} Completeness model for point-source detections as a function of $g$ (blue) and $i$ (red) magnitude estimates for the CSST. The dashed and solid lines represent the single- and double-exposure modes respectively. \emph{Right:} Expected photometric uncertainties in the two bands for the one-exposure model (crosses) and fitted model (dashed lines). The two-exposure uncertainty model (full lines) corresponds to a shift of these models by 0.38\,magnitudes to account for the increased exposure time. 
    }
    \label{fig:comp_uncert}
\end{figure*}

\section{Mock Data}
\label{sec:mock data}

\subsection{The CSST Imaging Survey}

The CSST wide imaging survey has a sky coverage of 17,500 deg$^2$ (42.4\% of the whole sky), and 400 deg$^2$ of this area will be selected for the deep survey (see details in Tab.~\ref{tab:CSST}). The designed photometric system has seven broad-band filters, $NUV$, $u$, $g$, $r$, $i$, $z$, and $y$, that cover the full 255--1065 nm range, from the near-ultraviolet ($NUV$) to the near-infrared ($y$). The transmission curves of the filters are shown in Fig.~\ref{filter} and listed in Table~\ref{tab:CSST}. For the wide survey, the survey depth is reached with two 150-second exposures for the $u$, $g$, $r$, $i$, and $z$ bands, and four such exposures for the $NUV$ and $y$ bands. The final products will correspond to stacked images whose photometric catalogues will, ultimately, reach the following 5$\sigma$ point-source limiting magnitudes: 25.4 ($NUV$), 25.4 ($u$), 26.3 ($g$), 26.0 ($r$), 25.9 ($i$), 25.2 ($z$), and 24.4 ($y$) in the AB system (see Tab.~\ref{tab:CSST}; \eg\citealt{Cao2018}).

In this work, we estimate the completeness and photometric uncertainties of point-source detections as a function of magnitude based on the designed parameters of the CSST \citep{Zhan2011,Cao2018}. Firstly, we use the CSST Image Simulator to produce mock optical images of stars for a single 150-second exposure. The catalogue consists of stars brighter than $G$ = 21 from Gaia DR2 and fainter MW stars simulated using the \texttt{Galaxia} model \citep{Sharma2011}, a fast code for creating a synthetic survey of the MW (see more details in Sec.~\ref{subsec:MW}). We then perform source detection and photometric measurements on the mock images using \texttt{SExtractor} \citep{Bertin1996}. The completeness functions of point sources are shown in the left panel of Fig.~\ref{fig:comp_uncert}, with the results for a single-exposure shown as dashed lines for the $g$ and $i$ bands (blue and red, respectively). For the nominal double-exposures depth, we shift those models by 0.38 magnitudes (solid lines). In the single-exposure mode, the detection limits (defined as the magnitudes at which the completeness drops to 70\%) are $g_{\rm lim}=26.0$ and $i_{\rm lim}=25.5$. 
Note that we assume star and galaxy can be separated 100$\%$
for point source detections to the survey depth. The estimates of the successful rate of star galaxy separation using different models as well as its impact on the dwarf galaxy detection will be investigated in future work.

The photometric uncertainties in each band are estimated from the same images. We first measure the magnitude of a detected object by processing mock images with SExtractor, then compare it with the true magnitude from the input catalogue that is given to the CSST Image Simulator. The uncertainty is described using the difference between the measured and the input values as a function of magnitude (see right panel of Fig.~\ref{fig:comp_uncert}). In the single-exposure mode, uncertainties reach $\sim 0.2$\,mag around $g = 25.5$ and $i = 24.7$. We model the measured uncertainties with the following model:

\begin{equation}
\centering
 \delta (M)=a+{\rm exp}(b M-c),
\label{eq:uncertainty}
\end{equation}

with $\delta(M)$ the uncertainty, $M$ the magnitude, and $(a,b,c)$ the coefficients we fit for. Based on the measured uncertainties, we find $(a,b,c)=(0.015, 0.736, 20.423)$ and $(0.026, 0.827, 22.144)$ for the $g$ and $i$ bands, respectively. The resulting fits are shown in Figure~\ref{fig:comp_uncert}. These are for a single CSST image so, to account for the double-exposure mode that will be used for these two bands, we simply shift those models by 0.38\,magnitudes.



\begin{figure*}
   \centering
   \includegraphics[width=19cm, angle=0]{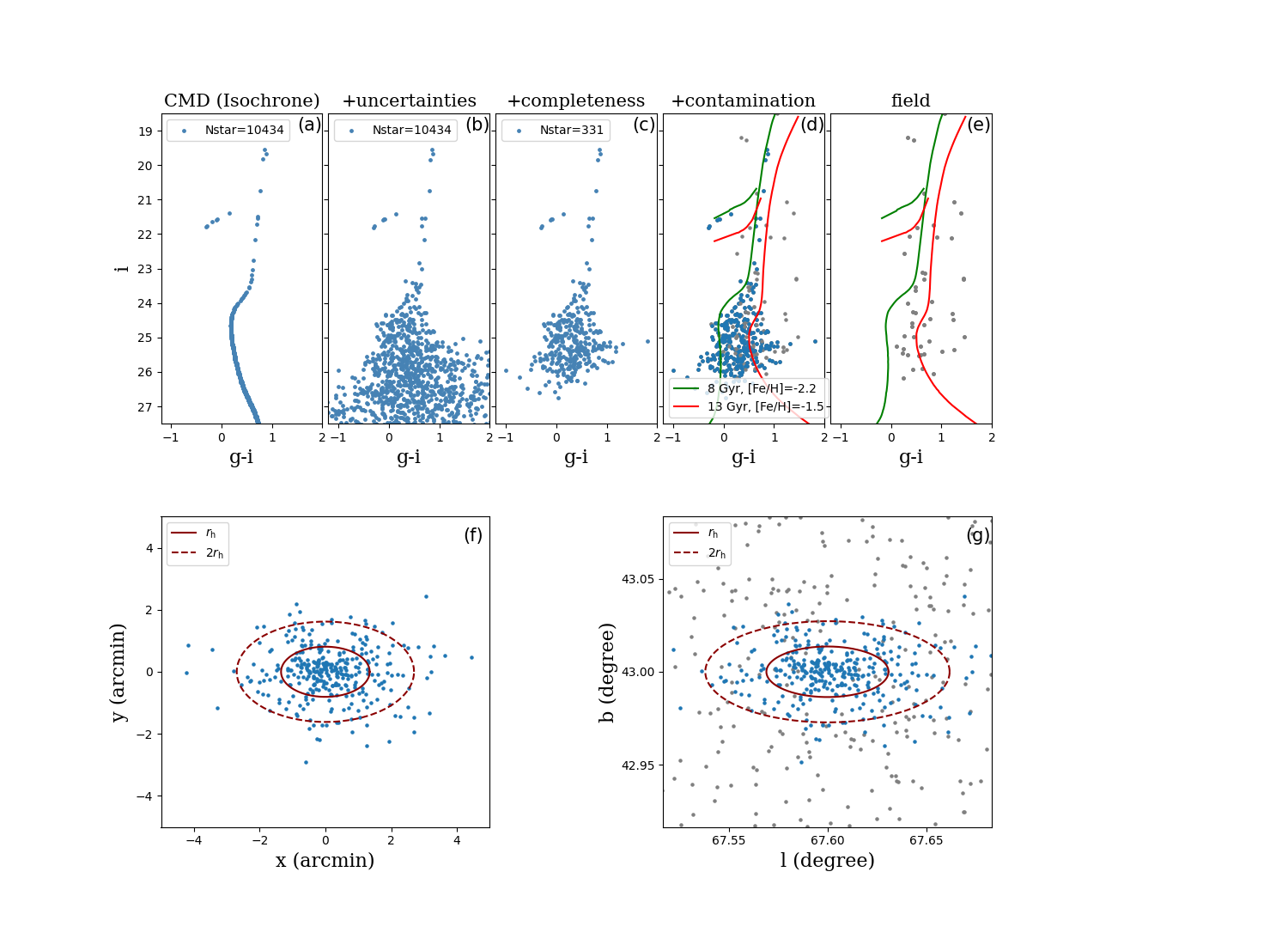}  
   \caption{
   Process of generating an artificial dwarf galaxy in the CMD (a)--(e) and on the sky (f)(g). Panel (a): a sample of 10434 stars drawn from the isochrone of a single stellar population ($M_{V}=-4.5$, [Fe/H] $=-2.2$, age = 10 Gyr). Panel (b): same as (a) after taking into account the photometric uncertainties. Panel (c): same (b) after folding in the completeness model. Panel (d): the final CMD of the artificial dwarf galaxy within $2r_{\rm h}$, combined with MW foreground stars over the same region (gray). The latter are plotted separately in panel (e) for comparison. The green and red lines represent the isochrones used to define the CMD selection box use by the search algorithm. Panel (f): the 2D spatial distribution in the Cartesian coordinates of stars from a dwarf galaxy with $r_{\rm h}=100\pc$ at the distance of 255 $\kpc$. The ellipsis correspond to 1 and 2 $r_{\rm h}$ (solid and dashed lines, respectiely). Panel (g): the stars from (f) after their ingestion in the median latitude MW field.
   }
   \label{mock-dg}
\end{figure*}

\subsection{Mock Catalogue }
\label{sect:data}

With the completeness and photometric uncertainty functions of point-source detection in hand, we now construct mock catalogues in the CSST bands by injecting artificial dwarf galaxies in the foreground MW model. In this work, we aim to explore the  detection efficiency of dwarf galaxies with CSST from a series of mock catalogues by varying five key parameters: the absolute magnitude ($M_V$, distance ($D$), and the half-light radius ($r_\mathrm{h}$) of an artificial dwarf galaxy, the Milky Way foreground parametrized by the Galactic latitude $b$, and the survey depth with single and double exposures. 

\subsubsection{Milky Way foreground}
\label{subsec:MW}
The mock MW foreground catalogue is generated by \texttt{Galaxia}. It uses a Galactic model that is initially based on the Besan\c{c}on model \citep{Robin2003}, as well as realistic substructure of the stellar halo based on N-body simulations \citep{Bullock2005}. The MW model used in this work includes a metal-rich thick disc correction with [Fe/H] from $-$0.78 to $-$0.16 according to \citet{Sharma2019}. 

Since the MW foreground is more crowded close to the Galactic disk, we generate three different MW fields that aim to represent the varying contaminating conditions in the CSST survey. Labeled "L", "M", and "H", they represent large sky areas centered on location at low ($b=25\deg$), median ($b=45\deg$) and high ($b=70\deg$) Galactic latitudes (see details in Table.~\ref{tab:mock data}).

\subsubsection{Artificial Dwarf Galaxy}
Most of the known dwarf galaxies in the Local Group are old and very metal-poor because of their ancient and relatively short star formation histories. In this work, we generate a group of artificial dwarf galaxies that host a single stellar population randomly drawn from the age range 8--13 Gyr. Taking into account the mass-metallicity relationship observed from the satellite dwarf galaxies \citep[\eg][]{Kirby2009}, we set a range of metallicities $-1.7<\FeH<-1.5$ for dwarf galaxies with $M_{V}\lesssim-7$ (stellar mass $\gtrsim$ 2$\times$10$^5$ \msun), and $-2.2<\FeH<-1.7$ for dwarf galaxies with $M_{V}\gtrsim-7$ (stellar mass $\lesssim$ 2$\times$10$^5$ \msun). With the age and metallicity of a dwarf galaxy chosen, the stellar parameters and luminosity function of this population are generated by the PAdova and Trieste Stellar Evolution Code (PARSEC) library \citep{Bressan2012}\footnote{\url{http://stev.oapd.inaf.it/cgi-bin/cmd}}. We then use \texttt{GalevNB} to generate CSST photometric bands based on these stellar parameters by convolving theoretical spectra with the response function of each filter \citep{Pang2016}.

The artificial dwarf galaxies are assumed to follow a typical projected radial density profile \citep{am2022PAndAS}, characterized by the half-light radius $r_{\rm h}$ of the system. The probability density function of stars as a function of the on-sky radius, $P(r)$, is defined as

\begin{equation}
        P(r)=\frac{1.68^{2}r}{2 \pi r_{\rm h}^2}{\rm exp}\left(-1.68\frac{r}{r_{\rm h}}\right).
\end{equation}

\noindent We further flatten this distribution so it has ellipticity $e=1-b/a$, with $a$ and $b$ the major and minor axis scale lengths, respectively.

The process to generate an artificial dwarf galaxy is visualized in Fig.~\ref{mock-dg} for $M_V=-4.5$, $\FeH=-2.2$, age=10\,Gyr, $D=255\kpc$, at high Galactic latitude ($b=43\deg$), and for the double-exposure mode. Panels (a) to (e) show how the color-magnitude diagram (CMD) of the system in constructed and panels (f) and (g) show its on-sky projection. Stars are first drawn from the galaxy's isochrone and corresponding luminosity function of chosen age, metallicity, and distance until the absolute magnitude of the system reaches the target $M_V$ (a). The CMD is then convolved by the photometric uncertainties (b) and the completeness model of CSST is applied (c). In (d), we combine the dwarf galaxy stars (blue) and the Milky Way foreground stars (grey, see also panel (e)) within $2r_{\rm h}$ at $(l,b)=(67.6\deg,43\deg)$. The spatial distribution of the dwarf galaxy stars are shown in panel (f) in coordinates of the plane tangent to the sky, and as they would appear in the data in panel (g).

Following this procedure, we construct a large series of mock dwarf galaxies in the range of $-10.0<M_{V}<+2.0$ with steps of 0.5, and $3\pc<r_{\rm h}<3\kpc$ with steps of 0.25 $\log(r_{\rm h}/pc)$. Distances are set within the range $32<D<2048\kpc$ (see details in Table~\ref{tab:mock data}) and the ellipticity of the dwarf galaxy is randomly sampled between $0.3<e<0.7$.

Mock catalogues are generated for the six different combinations of the three MW fields (L, M, H) and the two survey depths (single and double exposures). For each combination of these parameters, we generated 10,140 mock dwarf galaxies with three varying parameters ($M_V, D, r_\mathrm{h}$), and ingested them in the MW contamination catalogue in 20 different batches of $\sim500$\,galaxies each to avoid an overlap between the galaxies. In total, we simulate 60,840 artificial galaxies.


\section{Search algorithm}
\label{sec:method}

\begin{table*} 
	\caption{Chosen parameter ranges for the artificial dwarf galaxies and detection thresholds.}
	\setlength{\tabcolsep}{36pt}
	\label{tab:mock data}
	\centering  
	\begin{tabular}{l|lll} 
		\hline
		\hline
		\multicolumn{4}{c}{artificial dwarf\ galaxy}\\
		\hline
		Parameter& Minimal & Maximal  &Step (log scale)\\
		\hline
		$M_{V}$     & -10.0  & 2.0    & 0.5 \\
        $r_{\rm h}$ (pc)     & 3.16  & 3162    & 0.25\\
        D (kpc) & 32 & 2048    &   1 \\
		\hline
	\end{tabular}
	\setlength{\tabcolsep}{15pt}
	\begin{tabular}{ccc|c|cccc } 
    \multicolumn{8}{c}{detection threshold of each field}\\
	\hline  
		field & l ($\deg$) & b ($\deg$) &  & depth& $\sigma'=$2 & $\sigma'=$4 & $\sigma'=$8\\  
		\hline  
		L (low)   & 35$-$85 & 15$-$35  & &\makecell{ single\ exp\\\hline double\ exp}    
		   &\makecell{6.2\\6.6}&\makecell{ 5.5\\5.3}&\makecell{5.2\\5.0}\\ 
		\cline{1-3} \cline{5-8}
		M (median)   & 40$-$100 & 30$-$60  &\makecell{ S/N$_{\rm thr}$}  &\makecell{single\ exp\\\hline double\ exp}    
		   &\makecell{6.5\\6.1}&\makecell{ 5.6\\5.0}&\makecell{4.7\\5.0}\\ 
		 \cline{1-3} \cline{5-8}
		H (high)   & 10$-$160 & 60$-$80  & &\makecell{single\ exp\\\hline double\ exp}    
		   &\makecell{6.3\\6.3}&\makecell{ 5.3\\5.2}&\makecell{5.7\\5.4}\\    
		\hline
	\end{tabular}
\end{table*}

The dwarf galaxy search algorithm used in this work was described in \citet{Walsh2009} and has been widely adopted, for instance to explore the data from the SDSS \citep{Walsh2007,Koposov2008}, the DES \citep{Bechtol2015, Drlica-Wagner2015, Koposov2015}, Pan-STARRS \citep{Laevens2015a, Laevens2015b}, Hyper Suprime-Cam Subaru Strategic Program \citep{Homma2018}, and the Wide component of the DECam Local Volume Exploration Survey \citep{Mau2020}. Following the approach of \citet{Koposov2008}, \citet{Walsh2009}, \citet{Bechtol2015}, and \citet{Homma2019}, our version of the algorithm works as follows. For a given distance choice, we start by selecting stars that fall within a CMD selection box constructed from isochrones that represent the dominant stellar populations observed in dwarf galaxies. Specifically, we choose two single populations with age = 8\,Gyr, $\FeH=-2.2$ and age = 13\,Gyr, $\FeH=-1.5$, shown as the green and red lines in Fig.~\ref{mock-dg}, panels (d) and (e). The two isochrones are shifted left (blue) and right (red) by 0.05 mag along the $g-i$ color axis to form the contours, which are further widened by taking into account the photometric uncertainties. Stars located within the selection box are used to build binned density maps on the sky, $I(i)$, using \texttt{Healpix} \citep{Gorski2005}. These density maps of stars on the sky are then convolved with a Gaussian kernel $g(i,\sigma)$ of dispersion $\sigma$ to produce smoothed maps

\begin{equation}
    \centering
        L(i,\sigma)=I(i)\otimes g(i,\sigma).  
    \label{eq:smooth}
\end{equation}

An efficient way to optimize the search for dwarf galaxies is to perform this convolution with a kernel of the size $\sigma_2$ of the signal one is looking for, and subtract from it a map convolved by a much larger kernel, $\sigma_1$. The latter map will account for a smoothly varying contamination and optimize the resulting signal-to-noise (S/N). The resulting signal map is therefore

\begin{equation}
    \centering
        \Delta L(i)_{\sigma_{1},\sigma_{2}}=L(i,\sigma_{2})-L(i,\sigma_{1}).
    \label{eq:reduce background}
\end{equation}

Following \citet{Koposov2008}, we use three sets of ($\sigma_{1}$, $\sigma_{2}$) kernels: (28$'$, 2$'$); (28$'$, 4$'$); and (56$'$, 8$'$). The three different $\sigma_{2}$ correspond to the angular sizes of a dwarf galaxy with $r_{\rm h}=100{\pc}$ at different distances (\ie $40\kpc$, $100\kpc$, and $200\kpc$). We tailor the background kernel $\sigma_1$ such that it is at least 7 times larger than $\sigma_{2}$ to optimize the computing time. For the same reason, we bin the starting map $I$ using $0.5'$ pixels when $\sigma_{2}=2'$ or $4'$ and we use $2.0'$ pixels when using the larger kernels.

To evaluate the significance of local overdensities, we define the standard deviation $s(i)$ at the $i$-th pixel in the differential density map $\Delta L(i)$ as

\begin{equation}
        s(i)^2= \langle|\Delta L(j)_{\sigma_{1},\sigma_{2}}|^2\rangle_{r_{1},r_{2}}-|\langle\Delta L(j)_{\sigma_{1},\sigma_{2}}\rangle_{r_{1},r_{2}}|^2,
    \label{eq:local-variance}
\end{equation}

with $\langle\,.\,\rangle_{r_{1},r_{2}}$ the average over pixels within a ring around the $i$-th pixel between $r_1=0.5\deg$ and $r_2=1\deg$. The standard deviation $s(i)$ of the differential density map evaluates the Poisson noise for the $i$-th pixel, and thus the significance of the overdensity is defined as

\begin{equation}
S/N(i) = \Delta L(i)/s(i).
\end{equation}

To systematically search for the stellar overdensities resulting from dwarf galaxies with a variety of physical scales and distances, we follow the procedure described above for different distance assumptions by shifting the CMD selection box with distance moduli between 18.0 ($40\kpc$) and 26.5 ($2\Mpc$), with steps of 0.5\,magnitudes. After convolving these maps with three different kernel sets, we obtain $18\times3 = 54\,S/N$ maps.

These different maps have different noise properties. To evaluate these $S/N$ maps on the same scale, we stack the 18 maps with the same sets of kernels but at different distances by simply taking their average, and obtain three ``master $S/N$ map''. This step is also applied to the empty fields when determining the detection threshold in Sec. \ref{subsect:thres}. We then re-scale each master map to generate the final master maps, $S/N_{\rm r}$, by subtracting the mean pixel value of each map, $\langle S/N \rangle$, and normalizing with their standard deviation, $\sigma_{S/N}$, \ie

\begin{equation}
       S/N_{\rm r}(i)=\frac{S/N(i)-\langle S/N \rangle}{\sigma_{S/N}}.
    \label{eq:rescale}
\end{equation}

Each pixel in these three master maps, corresponding to the three choices of kernels, now expressed the significance of a detection and we search for pixels with high $S/N_{r}$ in each of the three master maps as candidate detections of dwarf galaxies.

\section{Application to the Mock Data}
\label{sec:appli}
\begin{figure*}
   \centering
   \includegraphics[width=15.0cm, angle=0]{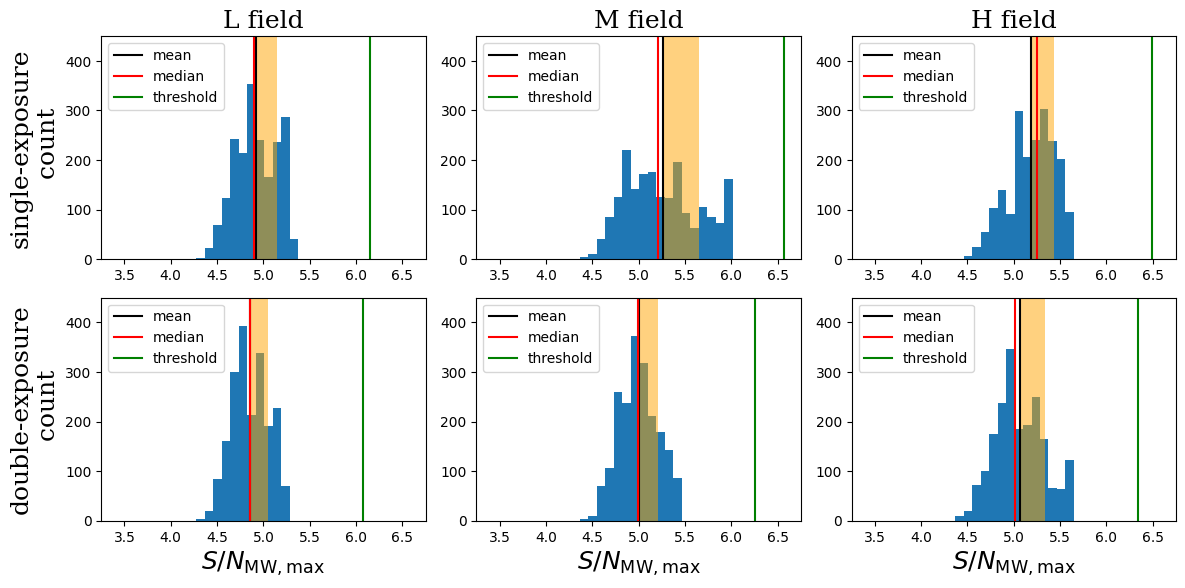} 
   \caption{Distribution of S/N$_{\rm MW,max}$ from 2000 S/N$_{\rm MW}$ maps with the set of ($\sigma_{1}=28'$, $\sigma_{2}=2'$) in the L, M, and H fields. The black and red lines denote the mean and median values of S/N$_{\rm MW,max}$, respectively and the orange band represents the standard deviation of the distribution. The green line marks the position of S/N$_{\rm thr}$. }
   \label{fig:thre max}
\end{figure*}
\begin{figure*}  
   \centering
   \includegraphics[width=15.0cm, angle=0]{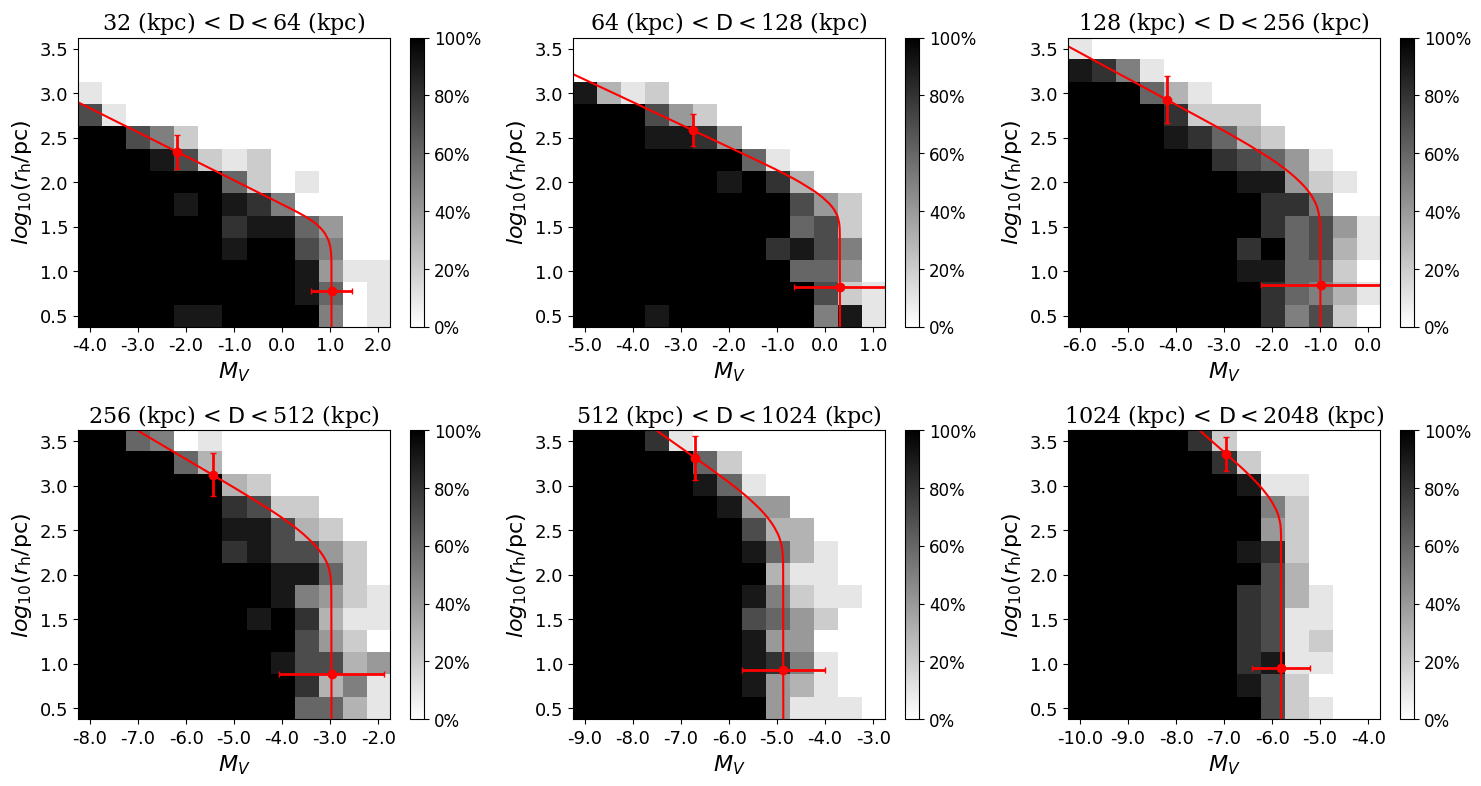}       
   \caption{Detection fraction map as a function of luminosity and size in different distance bins for M field, where the black pixels indicate 100\% detection fraction, and the white ones indicate 0\%. The red curve corresponds to the modeled  50\% recovery fraction and the error bars represent $\sigma_{M}$ and $\sigma_{r}$ as the widths of the transition regions.
   }
   \label{detect_m}
\end{figure*}

\subsection{Detection threshold}
\label{subsect:thres}
The application of the search algorithm described above to the mock catalogues results in three master maps, one for each set of kernels. We now need to decide on what constitutes a detection in those maps. It means determining the detection threshold above which we consider a detection reliable and significantly above noise fluctuations. This detection threshold, S/N$_{\rm thr}$, is calculated from the estimate of the Poisson fluctuations of maps generated from the mock MW foreground alone (\ie without the ingestion of any artificial dwarf galaxies), $S/N_{\rm MW}$ .

We first generate 2000 MW sky patches of 15\,deg$^{2}$ for the six combinations of Galactic latitudes (L, M, H) and exposure time (single- and double-exposure modes). 
We apply the same search algorithm from Sec. \ref{sec:method} to these empty fields for each distance moduli. For each patch, we first obtain 18 $S/N$ maps of MW foreground and then $S/N_{\rm MW, r}$ after stacking them. Fig.~\ref{fig:thre max} shows the distributions of the maximum values of these re-scaled maps, $S/N_{\rm MW, max}$ of 2000 patches in each combination, with the black vertical line denoting their mean value, $\langle$S/N$_{\rm MW, max}\rangle$. The stacking process averages the noises at different distances which reduces the effect from outliers and eventually leads to more successful detection rate. Since we did not add any artificial dwarf galaxy in these mocks, we use them to determine the threshold value of a detection, $S/N_{\rm thr}$, such that the number of false positives is below $1/500$, ensure that most of the detections correspond to meaningful objects without being overly conservative. Theoretically, this corresponds to setting $S/N_{\rm thr}=1.25\langle S/N_{\rm MW, max}\rangle$. For the 18000 $S/N_{\rm MW}$ we simulate, this corresponds to an expected 32 false detections and a low contamination rate of less than 0.17 \%. We are therefore confident that the chosen detection threshold does not bias our results and does not lead to detection limits that are too aggressive.

To verify our method, we applied the search algorithm to the public DES catalogs in a sky area of $\sim$14$^{\circ}\times13^{\circ}$ containing four new newly discovered dwarf galaxies (Tucana III, Tucana V, Phoenix 2, Tucana 2) \citep{Drlica-Wagner2015,Koposov2015}, out of which Phoenix 2 ($M_{V}=-2.8\pm0.2$ at 83 kpc) is close to the detection limit of DES at its distance. Following the above steps, all four galaxies have derived S/N above the detection thresholds of a M field listed in Tab. \ref{tab:mock data}.

\begin{figure*}
   \centering
   \includegraphics[width=15.0cm, angle=0]{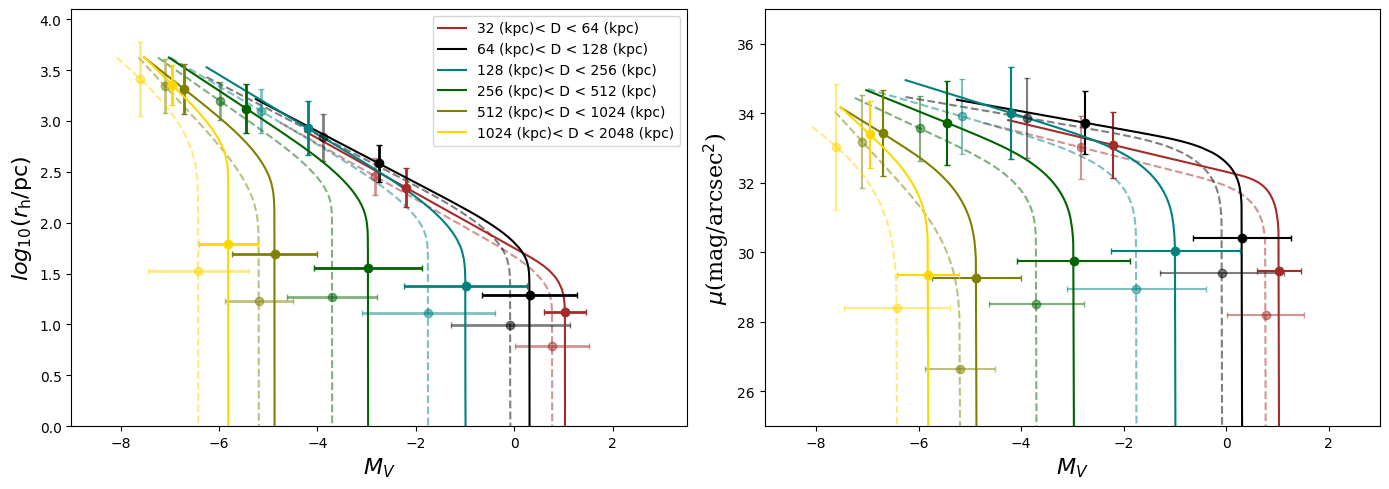}      
   \caption{Comparisons of detection limit as a function of $M_V$ and $r_{\rm h}$ in different distance bins for a M field. Same as Fig.~\ref{detect_m}, the lines represent the 50\% detection fraction, the error bars ($\sigma_{M}$, $\sigma_{r}$) represent difference between 16\% and 84\% detection fractions. The dashed and solid lines represent the results with single and double exposures respectively.   
   }
   \label{detect limit}
\end{figure*}

\begin{figure*}
   \centering
   \includegraphics[width=18.0cm, angle=0]{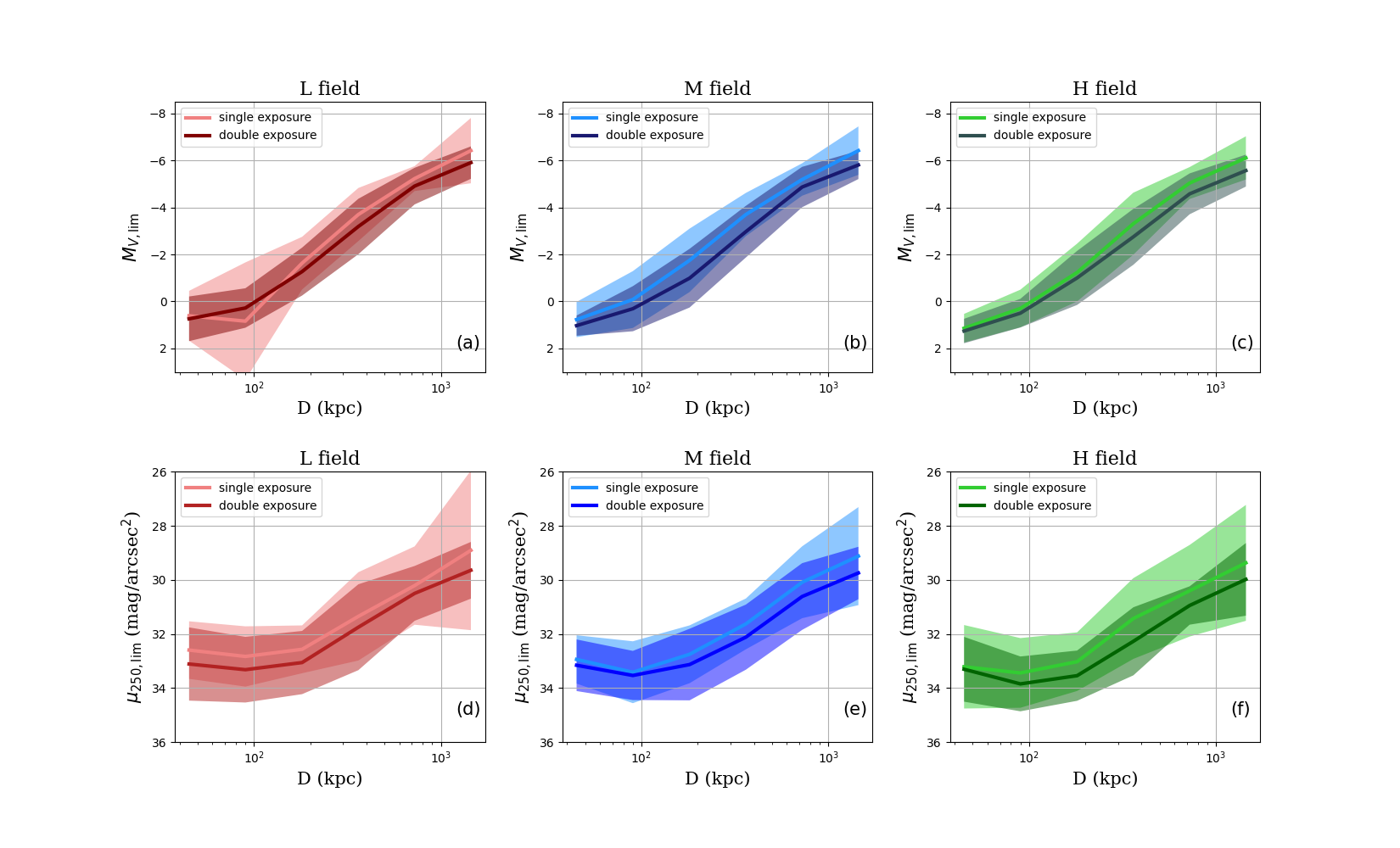}  
   \caption{Detection limit of $M_{V,\rm lim}$ as a function of distance for the two survey modes in the MW three fields. Top panels: results for the single- (light color) and double-exposure mode (dark color) in the L (panel a), M (panel b), and H (panel c) fields. The solid curve and the transparent band represent the mean trend and corresponding width ($\sigma_{M}$). Bottom panels: similar to the top panels but for the surface brightness limits of a galaxy with $r_{\rm h}$ = 250, $\mu_{250,\rm lim}$. The second exposure yields detection limits that can be up to $\sim$ 0.5 mag/arcsec$^2$ fainter.
   }
   \label{MV&u}
\end{figure*}

\begin{figure*}
   \centering
   \includegraphics[width=18.5cm, angle=0]{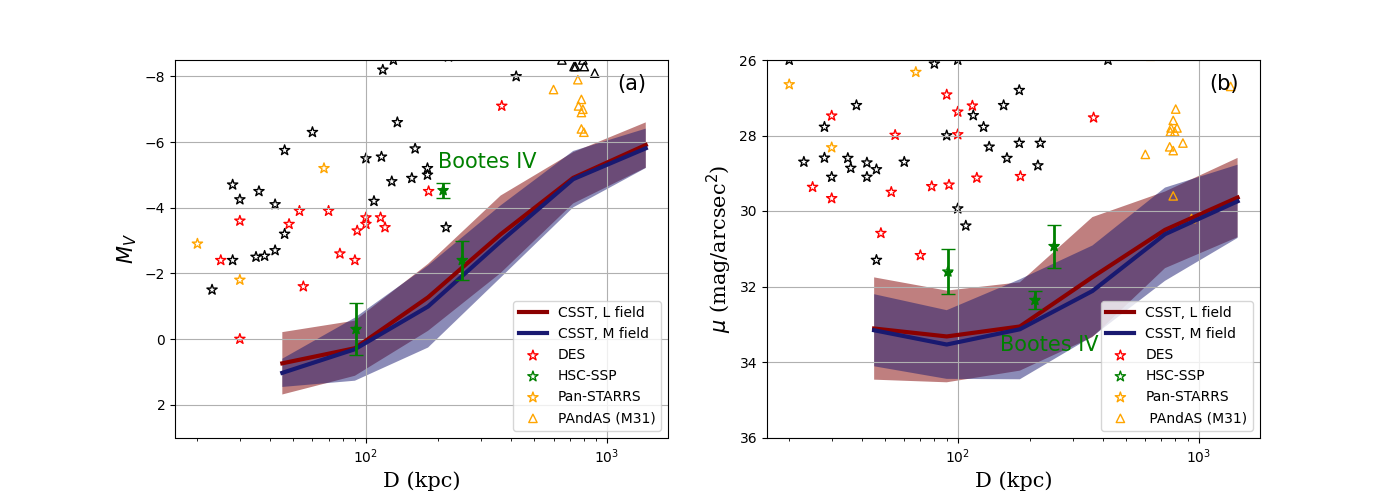}
   \caption{ Comparison of the detection limit of CSST with known Local Group satellite dwarf galaxies. Panel (a): The solid curves show the detection limits of $M_{V,\rm lim}$ as a function of distance in L (dark red) and M fields (dark blue) and the double-exposure mode, as previous shown in Fig.~\ref{detect limit}. The star and triangle symbols represent the satellites in the MW and M31, respectively, and are color-coded according to the survey in which they were discovered. Panel (b): Same as (a) but for the surface brightness limits, $\mu_{250,\rm lim}$ as a function of distance. In both panels, the three newly discovered dwarf galaxies from HSC-SSP are colored in green with error bars. Bo\"otes IV is located closest to the surface brightness detection limit in (b), but is the most luminous system of these three denoted in (a) as it is also the largest.
   }
   \label{compare}
\end{figure*}

\subsection{Results}

The resulting dwarf galaxy detection fractions for field M is shown in Fig.~\ref{detect_m}. The different panels represent separate distance bins spanning the range from 32$\kpc$ to 2$\Mpc$. The pixel size in the $(M_V$, $\log_{10}(r_{\rm h}))$ plane is (0.5, 0.25), and each pixel contains 10 galaxies. The gray levels represent the detection fraction in each pixel, with white and black pixels denoting 0\% and 100\% detection probabilities, respectively. Following previous work \citep[\eg][]{Koposov2008,am2022PAndAS}, we aim to fit an analytic model to these detection fractions to bypass the noise that is generated from testing only 10 dwarf galaxies per pixel. The transition region between full recovery and no recovery shows a ``knee'' feature: at faint magnitudes, the transition region is vertical for small half-light radii, before correlating with the radii at brighter magnitudes. We understand this behavior as follows. A minimum number of stars is required for a system to produce a detection for a given choice of detection threshold, irrespective of the size of this system. This effect produces the vertical part of the transition region with a fixed $M_V$ limit for the lower values of $\log(r_\mathrm{h})$. The sloped part of the transition region is more directly related to the surface brightness of the system and the recovery fraction in this region is therefore a function of both size and luminosity. These two different parts of the transition region also have different widths between the no and full detection modes.

We adopt an analytical model to characterize the recovery fractions as a function of $M_V$ and $\log_{10}(r_{\rm h})$, similarly to what was done by \citet{Koposov2008}, \citet{Drlica-Wagner2020}, or \citet{am2022PAndAS}. We parameterize the transition region with the following equation,

\begin{equation}  
    M_{V}=\begin{cases}
     M_{V, \rm lim} &\text{$r_{\rm h} < r_0$}\\
     \beta({\rm log}_{10}(r_0)-\log_{10}(r_{\rm h}))+M_{V,\rm lim} &\text{$r_{\rm h} > r_0$},
    \end{cases}
    \label{eq:fitting3}
\end{equation}
with $M_{V,\rm lim}$ the value of $M_V$ for the vertical part of the model, $r_0$ the half-light radius of the knee, and $\beta$ the slope of the sloped part of the model.

To model the width of the transition region, we further introduce $F(x)={\rm erfc}(x/\sqrt{2})/2$ and define the model recovery fraction as

\begin{equation}
\begin{aligned}
        f(M_{V},{\rm log}_{10}(r_{\rm h}))&=F\left(\frac{\Delta M_{V}}{\sigma_{M}}\right)F\left(\frac{\Delta \log_{10}(r_{\rm h}))}{\sigma_r}\right).    
    \label{eq:fitting0}
\end{aligned}
\end{equation}
Here, $\Delta M_{V}$ is the distance along the $M_V$ axis from the transition model of equation~\ref{eq:fitting3} for $\log_{10}(r_{\rm h})<\log_{10}(r_0)$ and $\Delta \log_{10}(r_{\rm h})$ is the distance along the $\log(r_\mathrm{h})$ axis for $\log_{10}(r_{\rm h})>\log_{10}(r_0)$. $\sigma_{M}$ and $\sigma_{r}$ represent the width of the transition regions for these two parts of the recovery fraction space, respectively.

With these definitions, the model has 5 parameters, ($M_{V,\rm lim}$, ${\rm log}_{10}(r_{0})$, $\beta$, $\sigma_{M}$, $\sigma_{r}$), which we fit for in all the recovery fraction spaces that we have calculated by minimizing residuals between the model and the data in the 5-dimension space. The red solid lines in each panel of Fig. \ref{detect_m} show the line of $50\%$ recovery fraction for the best fits and the error bars ($\sigma_{M}$ and $\sigma_{r}$) represent the interval between 16\% and 84\% detection fractions. The parameters for best fit model are listed in Tab.~\ref{tab:model param}--~\ref{tab:model param h} in the Appendix.

\begin{table} 
	\caption{Detection limits for field M and the double-exposure mode.}
	\setlength{\tabcolsep}{16pt}
	\label{tab:limit}
	\centering  
	\begin{tabular}{lcc} 
		\hline
		$D$ (kpc)& $M_{V, \rm lim}$ & $\mu_{250,\rm lim}$ (mag/arcsec$^2$)\\
		\hline
		32--64     & 1.0  & 33.2\\
        64--128     & 0.3  & 33.5\\
        128--256    & $-$1.0    & 33.1 \\
        256--512    & $-$3.0    & 32.1 \\
        512--1024    & $-$4.9    & 30.6 \\
        1024--2048    & $-$5.8    & 29.7 \\
		\hline
	\end{tabular}
\end{table}

Fig.~\ref{detect limit} summarizes the recovery fraction models in the size-luminosity and surface brightness-luminosity planes, for field M in different distance bins, and for the single- (dashed lines) and double-exposure models (full lines). The horizontal and vertical error bars denote the width of the transition region between full and no detection, \ie $\sigma_{M}$ and $\sigma_{r}$ from Eq. \ref{eq:fitting0}. We note that, as pointed out by \citet{am2022PAndAS}, that the detection limits do not follow lines of constant surface brightness, even for brighter luminosity than the knee (right panel), and should warn us against using simple surface-brightness cuts to summarize dwarf galaxy detection limits. It is likely the consequence of the complex search algorithm and is a common feature of such procedures \citet[\eg][]{Drlica-Wagner2020}.

We obtain similar results for the other two MW fields, L and H, and Figure~\ref{MV&u} presents a summary of the detection limits as a function of distance for the three MW fields and the two exposure configurations. In the top row of panels, $M_{V,\rm lim}$ corresponds to the vertical part of the models presented in Figures~\ref{detect_m} and~\ref{detect limit}. In the bottom row of panels, $\mu_{250,\rm lim}$ is the value of the detection limit model at a fixed radius\footnote{As mentioned above, the detection limits do not correspond to constant surface brightness limits, hence the need to assume the size of a fairly typical Local Group dwarf galaxy.} $\log_{10}(r_\mathrm{h})=250 \pc$. In all panel, the lighter/darker colored curve represents the detection limit for the single/double-exposure mode. The shaded bands denote the width of the transition region that was represented by error bars in Fig.~\ref{detect limit}. As expected, changing integration times from 150 to 300\,s yields a small but clear improvement to the detection limits, shifting them to $\sim0.5$\, magnitudes fainter. There is also a small impact of the MW contamination, with $M_{V,\rm lim}$ moving to slightly fainter magnitudes as the Galactic latitude increases. Similarly, $\mu_{250,\rm lim}$ only marginally changes with latitude. The artificial dwarf galaxies generated in this study are old and metal-poor. Although we expect to find more dwarf galaxies with recent star formation in the field of the Local Group, they are easier to be identified from their young blue stellar populations. Therefore, we quantify the detectability of the CSST using only the old dwarf galaxies. 


Fig.~\ref{compare} compares the predicted CSST detection limits with the properties of known Local Group satellites, including the lowest surface brightness systems that were only recently discovered in the deepest current survey, the Hyper Suprime-Cam Subaru Strategic Program (HSC-SSP). These three satellites, Virgo I, Cetus III, Bo\"otes IV \citep{Homma2016,Homma2018,Miyazaki2012} located at $D=91\kpc$,  $251\kpc$, and $209\kpc$ and with $M_V = -0.3$, $-2.4$, and $-4.5$ are the typical kind of the systems that the CSST survey should be able to uncover easily. These satellites are about 2 magnitudes fainter than the faintest dwarf galaxies discovered in the previous generation of surveys (SDSS, DES, Pan-STARRS). Even in the case of M31, whose satellite dwarf galaxies were mainly discovered in the data from the PAndAS survey that is deeper than these other surveys, CSST should be able to uncover new systems that are fainter than the currently known ones, further populating the faint end of M31's dwarf galaxy system.

For a more quantitative comparison, the detection limits of $M_{V, \rm lim}$ and $\mu_{250,\rm lim}$ are summarized in Tab.~\ref{tab:limit}, with the assumption of double exposures in the M field, at different distances. In particular, if we compare those with the DES and Pan-STARRS detection limits \citet{Drlica-Wagner2020}, we find that, at the distance of 400$\kpc$, the detection limits of these surveys that led to the discovery of numerous satellites are $M_{V,\rm lim} = -4.5$ (DES, equivalent to $\sim6\times10^3$\msun) and $-6.5$ (PS1), while the CSST survey should yield $M_{V,\rm lim} = -3.0$ (equivalent to $\sim1.6\times10^3$\msun). At this distance, the surface brightness limits reach $\mu_{250,\rm lim} = 29.9$ mag/arcsec$^{2}$ (DES) and 28.1 mag/arcsec$^{2}$ (PS1), compared to $\mu_{250,\rm lim} = 32.0$ mag/arcsec$^{2}$ for CSST. In the case of the M31 satellites, we use our results for field L and 512--1024\kpc\ results (M31 is located at $b=-21.6\deg$ and $D\sim780\kpc$) for comparison with the detailed study of \citet{am2022PAndAS} for PAndAS. They find $\mu_{250,\rm lim}\ge30.0$ mag/arcsec$^{2}$ in regions without much contamination from M31's stellar halo substructure, while our analysis yields a slight improvement with $\mu_{250,\rm lim}\ge30.5$ mag/arcsec$^2$.

Compared to other space-based telescope programs, \eg, Euclid has similar survey depths, 26.2 for the visible band and 24.5 for near-infrared bands \citep{Euclid&2022}. Roman is designed to reach almost 27 for infrared bands \citep{Akeson&2019}. The filter system design CSST has the narrowest bandwidth as well as the bluest wavelength coverage. Compared the ground-based telescope program, the LSST has the expected full depth of g $\sim$ 27.4 after 10 years of survey \citep{Ivezi'c2019}. The CSST is less deep (g $\sim$ 26.3) but has a better angular resolution, which is critical to separate stars and galaxies from faint sources. From all of the above, the designed CSST features are competitive to search for old and metal-poor star populations from dwarf galaxies. It will open the realm of efficient searches for dwarf galaxies beyond the Local Group and should, for instance, enable the discovery of very faint dwarf galaxies in the field. In the most distant bin (1 -- 2$\Mpc$) explored in this work, we predict that the detection limits are $M_{V,\rm lim} = -5.8$ and $\mu_{250,\rm lim}$ = 29.7 mag/arcsec$^2$ for the CSST.

\section{Summary}

\label{sec:sum}

In this work, we determine the detection limits of finding dwarf galaxies within $2\Mpc$ with the upcoming wide photometric survey that will be observed by the Chinese Space Station Telescope. We estimate the completeness of point-source detections and uncertainties in each photometric band and use these to construct mock catalogues in the CSST bands of artificial dwarf galaxies ingested in modeled MW foreground. From the application of a well-established search algorithm, we calculate the recovery fraction of dwarf galaxies in the ($M_{V}$, $r_{\rm h}$) space at different distances ranging from 32 kpc to 2 Mpc.

We show that the CSST is able to push the frontier of dwarf galaxy searches and detect fainter objects than currently known around the MW-M31 pair or, beyond, in the field. The surface brightness of most of the known MW satellites is brighter than 32.0 mag/arcsec$^{2}$, whereas the CSST survey will open the realm of fainter systems, down to $\mu_{250,\rm lim}\sim$ 33.0 mag/arcsec$^{2}$ within $\sim200\kpc$. Around M31, the detection limit can be slightly improved from 30.0 mag/arcsec$^{2}$ \citep[estimated for the PAndAS survey ][]{am2022PAndAS} to $\sim30.5$ mag/arcsec$^2$. In a future project, we will use the detection limits determined here to further predict the luminosity function we can hope to measure within the CSST volume by modeling observed satellite populations \citep{Koposov2008,Drlica-Wagner2020} from cosmological simulations \citep[\eg NIHAO][]{Wang2015}.

As we move further away from the Local Group, the CSST survey can still reach the detection limits of $M_{V,\rm lim}$ = $-$6.0, $\mu_{250,\rm lim}\sim29.5$ mag/arcsec$^{2}$ at the distance of 2$\Mpc$. It will provide us with the opportunity to discover many field dwarf galaxies expected beyond the sphere of influence of the Local Group. The planned survey has a wide coverage of 17,500 deg2 (42.4\% of the whole sky) that covers part of the northern sky, including M31, and overlaps with the Legacy Survey of Space and Time (LSST) in the South. From this joint effort in dwarf galaxy searches, we will, for the first time, have a sample of low-mass dwarf galaxies beyond the MW-M31 pair that can be directly compared to zoom-in dwarf galaxy simulations \citep[\eg EDGE][]{rey2019}.


\section*{Acknowledgements}
We acknowledge the science research grants from the China Manned Space Project with NO. CMS-CSST-2021-B03, the cosmology simulation database (CSD) in the National Basic Science Data Center (NBSDC) and its funding NBSDC-DB-10 (No.2020000088).

Z.Y., N.F.M. and R.A.I. acknowledge funding from the Agence Nationale de la Recherche (ANR project ANR-18-CE31-0017) and the European Research Council (ERC) under the European Unions Horizon 2020 research and innovation programme (grant agreement No. 834148).
X.K. acknowledge the support from the National Key Research and Development Program of China (No. 2022YFA1602903), the NSFC (No. 11825303), the China Manned Space project with NO.CMS-CSST-2021-A03, CMS-CSST-2021-B01

\section*{Data availability}
There are no new data associated with this article.

\bibliographystyle{mnras}
\bibliography{ref.bib} 




\appendix

\section{MODEL PARAMETERS}

\begin{table*}
	\centering
	\caption{model parameters of detectivity distribution in L field}
	\setlength{\tabcolsep}{16pt}
	\label{tab:model param}
	\begin{tabular}{llllll} 
		\hline
		D(kpc)&$M_{V,\rm lim}$& ${\rm log}_{10}(r_{0})$& $\beta$ &$\sigma_{M}$&$\sigma_{r}$\\
		\hline
		single exposure\\
		\hline
        32   --  64       &  0.61   &  1.45   &  0.27   &  1.06   &  0.21  \\
        64   --  128      &  0.84   &  1.42   &  0.28   &  2.5   &  0.22  \\
        128  --  256      &  -1.63   &  2.03   &  0.28   &  1.12   &  0.18  \\
        256  --  512      &  -3.7   &  2.34   &  0.32   &  1.13   &  0.33  \\
        512  --  1024     &  -5.23   &  2.53   &  0.39   &  0.53   &  0.29  \\
        1024 --  2048     &  -6.42   &  2.72   &  0.46   &  1.39   &  0.59  \\

		\hline
		double exposure\\
		\hline
		32 -- 64        & 0.74    & 1.58   & 0.26   & 0.95   & 0.27 \\
        64 -- 128       & 0.28    & 1.70    & 0.28   & 0.84   & 0.24 \\
        128 -- 256      & -1.27   & 2.05   & 0.29   & 1.02   & 0.23 \\
        256 -- 512      & -3.19   & 2.26   & 0.36   & 1.18   & 0.32 \\
        512 -- 1024     & -4.91   & 2.47   & 0.45   & 0.78   & 0.20 \\
        1024 -- 2048    & -5.91   & 2.82   & 0.41   & 0.69   & 0.21 \\
		\hline
	\end{tabular}
\end{table*}

\begin{table*}
	\centering
	\caption{model parameters of detectivity distribution in M field}
	\setlength{\tabcolsep}{16pt}
	\label{tab:model param m}
	\begin{tabular}{llllll} 
		\hline
		D(kpc)&$M_{V,\rm lim}$& ${\rm log}_{10}(r_{0})$& $\beta$ &$\sigma_{M}$&$\sigma_{r}$\\
		\hline
		single exposure\\
		\hline
		32   --  64       &  0.77   &  1.47   &  0.27   &  0.75   &  0.18  \\
        64   --  128      &  -0.08   &  1.88   &  0.25   &  1.21   &  0.23  \\
        128  --  256      &  -1.75   &  2.13   &  0.29   &  1.35   &  0.21  \\
        256  --  512      &  -3.70   &  2.43   &  0.34   &  0.92   &  0.19  \\
        512  --  1024     &  -5.19   &  2.35   &  0.52   &  0.69   &  0.27  \\
        1024 --  2048     &  -6.42   &  2.95   &  0.41   &  1.03   &  0.36  \\
		\hline
		double exposure\\
		\hline
		32 -- 64        & 1.03    & 1.47   & 0.27   & 0.43   & 0.19 \\
        64 -- 128       & 0.31    & 1.81   & 0.25   & 0.96   & 0.18 \\
        128 -- 256      & -0.99   & 1.99   & 0.29   & 1.25   & 0.27 \\
        256 -- 512      & -2.97   & 2.33   & 0.32   & 1.10    & 0.24 \\
        512 -- 1024     & -4.87   & 2.61   & 0.38   & 0.86   & 0.25 \\
        1024 -- 2048    & -5.81   & 2.81   & 0.48   & 0.60    & 0.19 \\
		\hline
	\end{tabular}
\end{table*}

\begin{table*}
	\centering
	\caption{model parameters of detectivity distribution in H field}
	\setlength{\tabcolsep}{16pt}
	\label{tab:model param h}
	\begin{tabular}{llllll} 
		\hline
		D(kpc)&$M_{V,\rm lim}$& ${\rm log}_{10}(r_{0})$& $\beta$ &$\sigma_{M}$&$\sigma_{r}$\\
		\hline
		single exposure\\
		\hline
		32   --  64       &  1.14   &  1.48   &  0.26   &  0.61   &  0.31  \\
        64   --  128      &  0.30   &  1.71   &  0.28   &  0.80   &  0.26  \\
        128  --  256      &  -1.23   &  2.01   &  0.31   &  1.23   &  0.22 \\
        256  --  512      &  -3.31   &  2.15   &  0.39   &  1.32   &  0.30  \\
        512  --  1024     &  -5.04   &  2.59   &  0.41   &  0.68   &  0.34  \\
        1024 --  2048     &  -6.11   &  2.78   &  0.44   &  0.92   &  0.43  \\
		\hline
		double exposure\\
		\hline
		32 -- 64        & 1.26    & 1.52   & 0.25   & 0.52   & 0.24 \\
        64 -- 128       & 0.50     & 1.83   & 0.26   & 0.61   & 0.20 \\
        128 -- 256      & -1.00    & 2.15   & 0.27   & 1.15   & 0.19 \\
        256 -- 512      & -2.75   & 2.30    & 0.33   & 1.19   & 0.25 \\
        512 -- 1024     & -4.58   & 2.57   & 0.38   & 0.87   & 0.14 \\
        1024 -- 2048    & -5.57   & 2.84   & 0.41   & 0.68   & 0.27 \\

		\hline
	\end{tabular}
\end{table*}

\begin{figure*}
   \centering
   \includegraphics[width=15.0cm, angle=0]{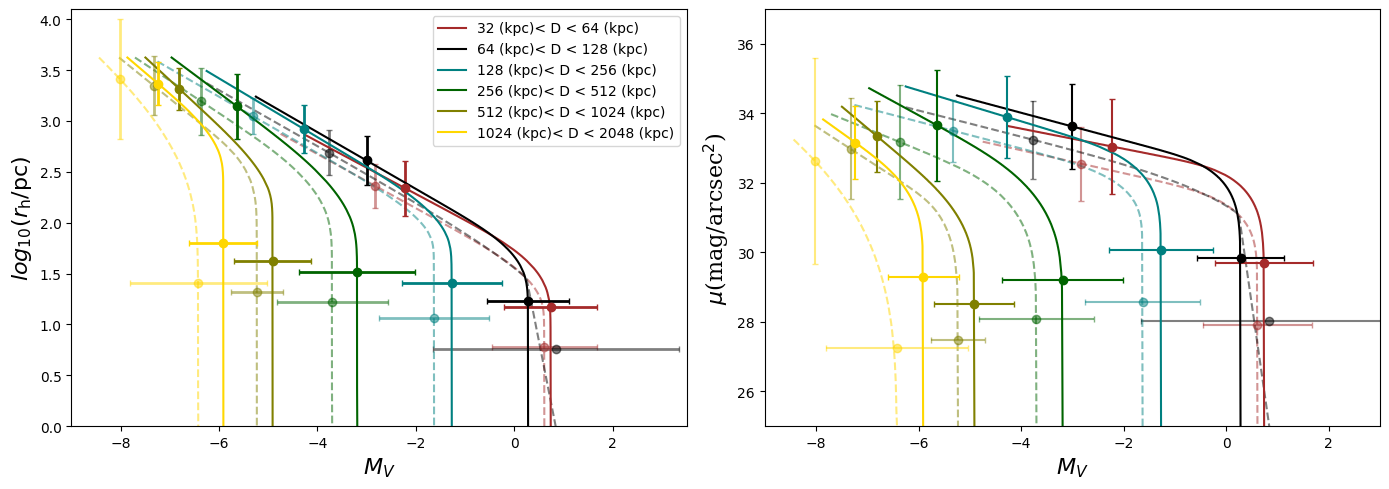}      
   \caption{Same as Figure~\ref{detect limit} but for MW field L.   
   }
   \label{detect limit l}
\end{figure*}

\begin{figure*}
   \centering
   \includegraphics[width=15.0cm, angle=0]{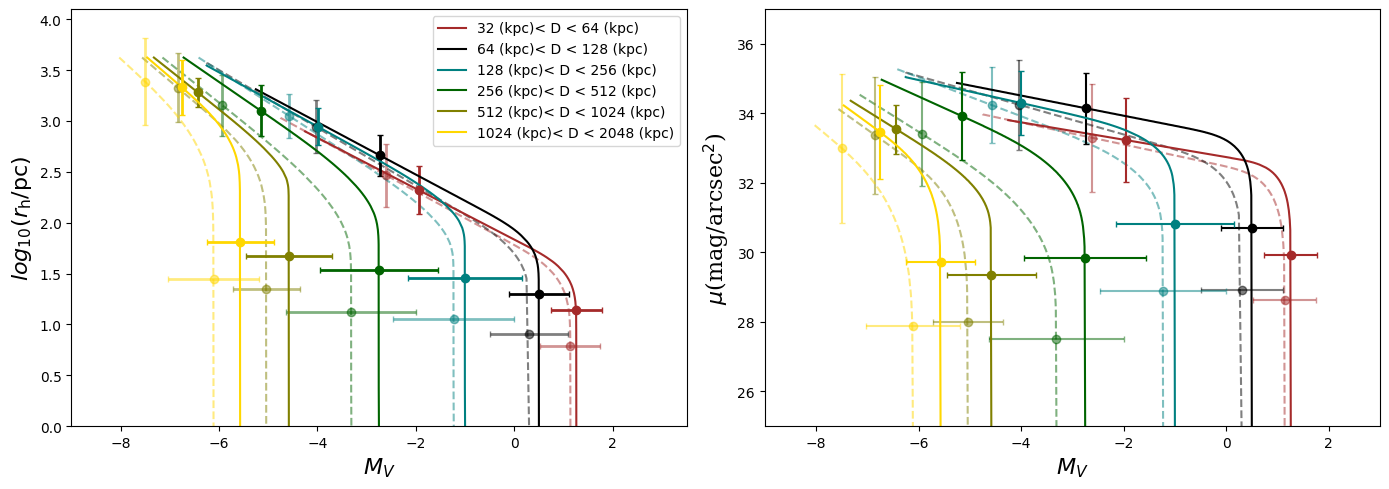}      
   \caption{Same as Figure~\ref{detect limit} but for MW field H.    
   }
   \label{detect limit h}
\end{figure*}

\bsp	
\label{lastpage}
\end{document}